# Spin injection and detection in lanthanum- and niobium-doped $SrTiO_3$ using the Hanle technique

Wei Han[1], Xin Jiang[1], Adam Kajdos[2], See-Hun Yang[1], Susanne Stemmer[2], Stuart Parkin[1†]

[1]IBM Almaden Research Center, San Jose, California 95120, USA

[2]Materials Department, University of California, Santa Barbara, California 93106-5050, USA

† e-mail: Stuart.Parkin@us.ibm.com

**There has been much interest in the injection and detection of spin polarized carriers in semiconductors for the purposes of developing novel spintronic devices. Here we report the electrical injection and detection of spin-polarized carriers into Nb-doped strontium titanate (STO) single crystals and La-doped STO epitaxial thin films using MgO tunnel barriers and the three-terminal Hanle technique. Spin lifetimes of up to ~100 ps are measured at room temperature and vary little as the temperature is decreased to low temperatures. However, the mobility of the STO has a strong temperature dependence. This behavior and the carrier doping dependence of the spin lifetime suggest that the spin lifetime is limited by spin-dependent scattering at the MgO/STO interfaces, perhaps related to the formation of doping induced $Ti^{3+}$. Our results reveal a severe limitation of the three-terminal Hanle technique for measuring spin lifetimes within the interior of the subject material.**

## Introduction

Interest in oxide thin films and heterostructures has been stimulated by potential nano-electronic applications especially since functional oxides can be integrated with mainstream Si based semiconductor electronic platforms[1-3]. Over the past decade, the properties of SrTiO$_3$ (STO) interfaces and thin films, in particular, have attracted much attention, stimulated especially by the discovery of a high mobility two dimensional electron gas (2DEG) at the interface between STO and several other perovskites[1,2,4], as well as in La-doped STO thin films[5]. The possibility of spin injection and accumulation in STO films and at the 2DEG interface between LaAlO$_3$/STO[6] promises novel spin-based devices.

Here, we report the successful injection of spin-polarized electrons and their accumulation in bulk Nb-doped STO substrates and La-doped STO thin films at temperatures as high as room temperature, as characterized via a three-terminal Hanle measurement[7-11]. The current flows from an ohmic contact (Al/Ta/Au) through the STO and back through a ferromagnetic (FM) electrode that acts as the spin injector / detector (MgO/CoFe/Ta/Au), as shown in Fig. 1a. The spin dependent chemical potential is characterized by the voltage between another ohmic contact as the reference and the FM injector/detector. The spin injection through a tunneling barrier (2 nm MgO) yields a net spin accumulation in the STO conduction band, which is described by the difference of the spin up and spin down electrochemical potentials ($\Delta\mu = \mu_\uparrow - \mu_\downarrow$), as shown in Fig. 1b. The spin accumulation is measured as $\Delta V = \gamma \Delta\mu / 2e$, where $\gamma$ is the tunneling spin polarization of the in-plane magnetized FM spin injector/detector (CoFe). The spin accumulation is detected by applying an out-of-plane magnetic field ($B_z$) perpendicular to the electron spin direction (Fig. 1c), which causes the spins to precess in-plane, resulting in a decrease of the measured spin accumulation due to spin dephasing. The spin precesses at the Larmor frequency defined by $\varpi_L = g\mu_B B/\hbar$, where $g$ is the g-factor (g = 2 for STO)[5], $\mu_B$ is the Bohr magneton, and $\hbar$ is the reduced Planck's constant.

## Results

### Spin injection into Nb-doped STO using the Hanle technique

The Hanle curve measures the voltage between the STO and CoFe as a function of out-of-plane magnetic field. A typical Hanle curve is shown in Fig. 1d, which is measured on a Nb-doped STO sample (labeled Nb:STO1; Nb doping: 0.7wt%, $1.2\times 10^{19}$ cm$^{-3}$) at 10 K with a DC current of 500 µA. The spin lifetime (τ) is about 75 ps, as obtained by fitting the measured results with the equation $\Delta V(B_z) = \Delta V(0)/(1+(\varpi_L \tau)^2)$ (red curve in fig. 1d). The measured spin signal $\Delta V_{3T}$ is characterized by the difference between the voltages measured without magnetic field and when the magnetic field is large enough to cause the net spin accumulation to approach zero. In contrast to the nonlocal Hanle measurement [12,13], the measured spin lifetime directly probes the spin properties in the STO beneath the CoFe injector/detector. Due to spin relaxation induced by random magnetic fields from, for example, the roughness of CoFe film, a lower bound of the spin lifetime is found[14]. To confirm the authenticity of the measured spin signal, we measure the inverted Hanle effect by applying an in-plane magnetic field to enhance the component of the effective field along the direction of the spin polarization of the injected electrons (Fig. 1e). At high magnetic fields, the Hanle and inverted Hanle curves match each other (see Supplementary Figure S1), which confirms the success of spin injection and detection in STO[10].

Whereas spin injection corresponds to electrons flowing from the CoFe electrode into the STO (positive DC current), spin extraction refers to the opposite flow of electrons from STO into the CoFe (negative DC current). The latter process also gives rise to spin accumulation in the STO[15] since majority electrons preferentially flow from STO to CoFe, thereby leaving behind minority spin polarized carriers in the STO. Thus, the difference between the majority and minority spin chemical potentials will be negative. Fig. 2a compares the DC bias current dependence of the spin signal with that of the voltage drop across the tunnel barrier for bulk Nb-doped STO (labeled Nb:STO 1). At positive DC bias (spin injection), the spin signal is higher than that for negative DC bias (spin extraction), indicating a higher spin accumulation in the former case. This behavior is likely related to the bias dependence of the electronic structure and /or the bias dependence of the spin injection/detection efficiency[16]. The bias dependence of the spin lifetime for this same sample was studied at 10 K, as summarized in Fig. 2b. The spin lifetime is ~100 ps and depends only weakly on the bias current. To confirm this result a second Nb doped STO sample (labeled Nb:STO 2) with nominally the same doping concentration (0.7wt%, $1.2\times 10^{19}$ cm$^{-3}$) as that of Nb:STO1 was also studied at 10 K. The bias dependence of the spin lifetime for these two samples, shown in Fig. 2b., are very similar.

**Temperature dependence of the spin singals.**

The temperature dependence of the Hanle curves measured on Nb:STO1 for spin injection (500 µA) and spin extraction (-500 µA) are shown in Fig. 3a and Fig. 3b, respectively. The asymmetry of the spin injection and extraction persists up to room temperature. At room temperature, the spin signals are ~ 0.3 mV and ~ -0.1 mV for 500 µA and -500 µA bias currents, respectively. The successful electrical spin injection and accumulation at room temperature augurs well for possible future applications of oxide spintronics. A summary of the spin signals as a function of temperature are plotted in Fig. 3c. The spin voltages vary from ~0.1 to ~0.7 mV and the corresponding spin resistance-area products, $R_s \times A$, where A is the area of the contact (125 × 170 µm$^2$), are between 6 and 40 kΩ.µm$^2$. We note that these are significantly higher than theoretical values calculated from our measured spin diffusion lengths that we estimate vary from ~0.01 to 0.5 kΩ.µm$^2$. Such a discrepancy has previously been observed in Hanle studies of spin injection into various semiconductors but its origin is poorly understood[18]. One of these mechanisms attributes the high $R_s \times A$ values (at a GaAs / Al$_2$O$_3$ interface)[17] to a two-step tunneling mechanism via localized states. Such states could be formed at the Ti$^{3+}$ magnetic centers in our devices.

**Temperature dependence of the spin lifetimes.**

The summary of the spin lifetimes as a function of temperature are plotted in Fig. 3c and Fig. 3d. Although the spin lifetimes show little variation as a function of temperature, it is very interesting that the mobility, by contrast, exhibits a very strong temperature dependence, as shown in Fig. 3e. Indeed, the mobility in the Nb:STO1 sample increases by about a hundred-fold from room temperature to 10 K.

**Spin injection into La-doped STO.**

High quality, epitaxial, stoichiometric La-doped STO films with high mobilities have been grown recently by oxide molecular beam epitaxy (MBE)[5]. Here, we focus on two different samples with carrier concentrations of 2.9 × 10$^{19}$ cm$^{-3}$ and 6.0 × 10$^{19}$ cm$^{-3}$ as determined by Hall bar measurements. For the first sample with the lower dopant concentration, the Hanle spin signal can be detected up to ~200 K. The spin lifetime is ~50 ps and shows both a weak bias dependence as well as a weak temperature dependence, as summarized in Fig. 4a and 4b, respectively. For the sample with a higher dopant concentration, the Hanle spin signal can be detected up to ~100 K. The spin lifetime is found to be ~ 20 ps and is much shorter compared to

`

the first sample with the lower dopant concentration. These results are summarized in Fig. 4d and Fig. 4e.

**Discussion**

An important observation is that in all of our samples the spin lifetime shows little variation with temperature whereas, by contrast, the mobility exhibits a strong temperature dependence (Fig. 4c and 4f). This observation strongly suggests that the spin lifetime is limited by spin-dependent scattering at the interfaces with the MgO tunnel barrier. Specifically, we hypothesize that the interface scattering is likely related to the formation of $Ti^{3+}$ magnetic centers at the MgO/STO interface formed by chemical doping of Nb or La[19]. Indeed, evidence for such scattering is found from the Kondo effect observed in 2DEGs formed at the surface of STO single crystals when electrolyte gated[20,21] and the observation of magnetism at the LAO/STO interface[22-24], which have both been attributed to the formation of $Ti^{3+}$ magnetic centers[20,21,25,26].

The spin lifetimes measured at 10 K on the various samples discussed above are summarized in Fig. 5a as a function of bias voltage. The pink color corresponds to a Nb doping level of $1.2 \times 10^{19}$ $cm^{-3}$, and the green and gray colors correspond to La doping levels of $2.3 \times 10^{19}$ $cm^{-3}$ and $6.0 \times 10^{19}$ $cm^{-3}$, respectively. The dependence of the spin lifetime on the carrier density and mobility are shown in Fig. 5b and 5c, respectively, for data corresponding to a bias voltage of ~ -0.8 V. The mobility of La doped STO decreases with increasing carrier concentration due to increased ionic impurity scattering[5,27]. The higher mobility of the La doped as compared to the Nb doped STO can be attributed to the very high structural quality of the STO films[5].

Of considerable interest, the spin lifetime decreases systematically with increasing dopant concentration (Fig. 5). We suggest that is due to a corresponding increase in the number of magnetic centers at the STO/MgO interface. Indeed previous x-ray photoemission spectroscopy studies on Nb-and La-doped STO[19] show that the Nb and La dopants have valence states of $Nb^{5+}$ and $La^{3+}$, respectively, which gives rise to $Ti^{3+}$ centers; and that the concentration of $Ti^{3+}$ scales with the dopant concentration. Since $Ti^{3+}$ has a magnetic moment we hypothesize that the spin lifetime is limited by scattering from the $Ti^{3+}$ located at or near the MgO/STO interfaces. We note that longer spin lifetimes in Nb doped STO as compared to La doped STO may also be attributed to a difference in spin-orbit coupling constants resulting from the smaller atomic number of Nb.

As mentioned earlier, the inverted Hanle and Hanle curves overlap each other at a magnetic field of ~2T: this is the field required to saturate the magnetization of the CoFe electrode, as seen, for example, in measurements of the anisotropic magnetoresistance of CoFe (see Supplementary Figure S1c). When the moment of the CoFe electrode is oriented along the field (above 2T) there can be no Hanle or inverted Hanle effect. Thus, our experiments are limited to fields below 2 T. However, very large magnetic fields are required to saturate the paramagnetic moments on the $Ti^{3+}$ (s =1/2) centers. For example, at 10 K, assuming the magnetic moments follow a Brillouin function with no exchange between the moments, a field of 2 T will orient the moments to only 13% of their saturation moment. When the temperature is higher the moments will be even less ordered by the magnetic fields in which we can observe a Hanle or inverted Hanle effect. Thus, we conclude that the $Ti^{3+}$ moments will be fluctuating significantly for all the temperatures and magnetic fields in which we observe the Hanle or inverted Hanle effects. We conclude that it is magnetic scattering from these randomly oriented moments that depolarize the injected carriers and that this mechanism is weakly dependent on temperature and therefore

`

consistent with the weak temperature dependence that we find in our Hanle and inverted Hanle data (see Supplementary Figure S3).

In summary, spin injection and spin detection in doped STO bulk crystals and thin films has been achieved up to room temperature using MgO tunnel spin injectors. The spin lifetimes are about 100 ps, which are similar to conventional semiconductors such as GaAs and Si with similar carrier densities and doping levels [8,10,11,17]. Interestingly, we find that the spin lifetime has a very weak temperature dependence even though the carrier mobility in the doped STO samples has, by contrast, a very strong temperature dependence. We thus conclude that the spin lifetime is limited by spin-dependent scattering from magnetic centers at the STO/MgO interfaces which we attribute to $Ti^{3+}$ formed as a result of the La and Nb doping. Our results show that the Hanle technique to measure the spin polarization of carriers injected via a tunnel contact is strongly influenced by the formation of these contacts and thus may not be a reliable indicator of the spin lifetime within the interior of the STO itself.

## Methods

**Device Fabrication**

Nb doped SrTiO3 (0.7 wt%) substrates were obtained from Princeton Scientific Corporation. La-doped $SrTiO_3$ films were epitaxially grown on undoped $SrTiO_3$ (001) substrates using oxide MBE, as reported in detail elsewhere[5]. The Nb:STO substrates and La:STO films were cleaned using the following procedure: wash in methanol in an ultrasonic bath for 10 mins, followed by a distilled water wash in an ultrasonic bath for 10 mins, a dip in Hydrogen peroxide for 30 sec, and finally a distilled water rinse. Then the samples were heated in a furnace at either 1000 C (Nb: STO) or 900 C (La:STO), for ~90 mins in a pure oxygen environment that results in the

`

formation of atomically flat surfaces, that we confirmed by atomic force microscopy[28]. The spin injection/ accumulation devices were fabricated using shadow masks in an ultra high vacuum sputtering system. The tunneling contact is formed from 2 nm MgO, 10 nm CoFe, 5 nm Ta, and 20 nm Au and has an area of $125 \times 170$ μm$^2$. The ohmic contact is formed from 10 nm Al, 5 nm Ta and 20 nm Au. The RMS roughness averaged over an area of $10\times10$ μm was ~0.16 nm for the as prepared Nb:STO1 sample, and, was increased only slightly to ~0.29 nm after a 2 nm thick MgO tunnel barrier was deposited on top of the same sample.

**Electrical Measurement**

The electrical measurements were performed in a PPMS (Quantum Design) system. The mobility, carrier density, and resistivity were characterized using van der Pauw and Hall bar geometries. The carrier density shows little temperature dependence for both Nb:STO single crystals (see Supplementary Figure S2) and La:STO films[5]. The carrier density values used here are those measured at room temperature.

**Contact Characterization**

The temperature dependent I-V characteristics of a typical contact (CoFe/MgO/Nb:STO1) are characterized by the three terminal method and shown in Fig. 6. The I-V characteristics are highly non-linear and the contact resistance shows a modest increase as the temperature decreases, which are two important characteristics of tunneling contacts[29].

**Author Contributions**

**W.H. and S. P. conceived the experiments. W. H. carried out the measurements and analyzed the data with the help of X.J.. A. K. and S. S. prepared the La-dopes STO films.**

`

W.H. and S.P. wrote the manuscript. All authors discussed the data and commented on the manuscript.

**Competing financial interests:**

The authors declare no competing financial interests.

**Acknowledgements**

We acknowledge partial support for this work from the MURI program of the Army Research Office (Grant No. W911-NF-09-1-0398) and the King Abdullah University of Science & Technology (KAUST), and thank Prof. Aurelien Manchon, KAUST, for useful discussions.

**Figure Legends:**

**Fig. 1. Device geometry and three-terminal Hanle measurement.** (a) Device geometry: the tunneling contact is formed from 2nm MgO/CoFe/Ta/Au. (b) Tunneling spin injection: the spin-dependent chemical potential is split for spin up and spin down carriers. (c) Hanle spin measurement schematic. (d) Hanle measurement on a Nb (0.7 wt%) doped $SrTiO_3$ single crystal (Nb:STO1) with a carrier doping of $1.2 \times 10^{19}$ cm$^{-3}$ (T = 10 K, $I_{DC}$ = 500 µA). The arrow indicates the Hanle spin voltage signal ($\Delta V_{3T}$). (e) Inverted Hanle and Hanle curves measured with an in-plane ($B_{//}$) and an out-of-plane magnetic field ($B_Z$), respectively (T = 10 K, $I_{DC}$ = 500 µA).

**Fig. 2. Bias dependence of the Hanle signal.** (a) Spin voltage signal $\Delta V_{3T}$ as a function of injection current (red squares), and, current-voltage characteristic of the $CoFe/MgO/SrTiO_3$ tunnel injector (black circles) for the Nb:STO1 sample at 10 K. (b) Bias dependence of the spin lifetime measured at 10 K on two Nb (0.7 wt%) doped $SrTiO_3$ single crystals (black squares for Nb:STO1 and black circles for Nb:STO2) with the same carrier doping of $1.2 \times 10^{19}$ cm$^{-3}$. Error bars define the standard errors obtained by the Lorentz fitting.

**Fig. 3. Temperature dependence of the Hanle signal.** (a-b) Hanle curves for Nb doped $SrTiO_3$ (Nb:STO1 sample) at various temperatures for positive and negative bias currents of 500 µA. Temperature dependence of (c) spin voltage signal (red squares for $I_{DC}$ = 500 µA and black squares for $I_{DC}$ = -500 µA), (d) spin lifetime (red squares for $I_{DC}$ = 500 µA and black squares for $I_{DC}$ = -500 µA), and (e) mobility for the same sample. Error bars define the standard errors obtained by the Lorentz fitting.

**Fig. 4. Bias and temperature dependence of the Hanle signal for La doped STO.** Bias dependence of the spin lifetime at 10 K and temperature dependence of the spin lifetime and mobility measured for two La-doped SrTiO$_3$ films with doping concentrations of (a-c) $2.9 \times 10^{19}$ cm$^{-3}$ and (d-f) $6.0 \times 10^{19}$ cm$^{-3}$. ($I_{DC}$ = - 5 mA in Fig. 4b and $I_{DC}$ = - 100 μA in Fig. 4b) Error bars define the standard errors obtained by the Lorentz fitting.

**Fig. 5. Carrier doping dependence of the spin lifetime in STO.** (a) Bias dependence of the spin lifetime measured for various dopant concentrations at 10 K (red squares/red open squares for Nb:STO1/Nb:STO2 with the same carrier doping of $1.2 \times 10^{19}$ cm$^{-3}$, green squares for La:STO with the carrier doping of $2.9 \times 10^{19}$ cm$^{-3}$ and black squares for La: STO with the carrier doping of $6.0 \times 10^{19}$ cm$^{-3}$). Spin lifetime as a function of (b) carrier density and (c) mobility measured at 10 K. Error bars define the standard errors obtained by the Lorentz fitting.

**Fig. 6. Temperature dependent I-V characteristics of a typical contact on a Nb doped STO device.**

Figure 1

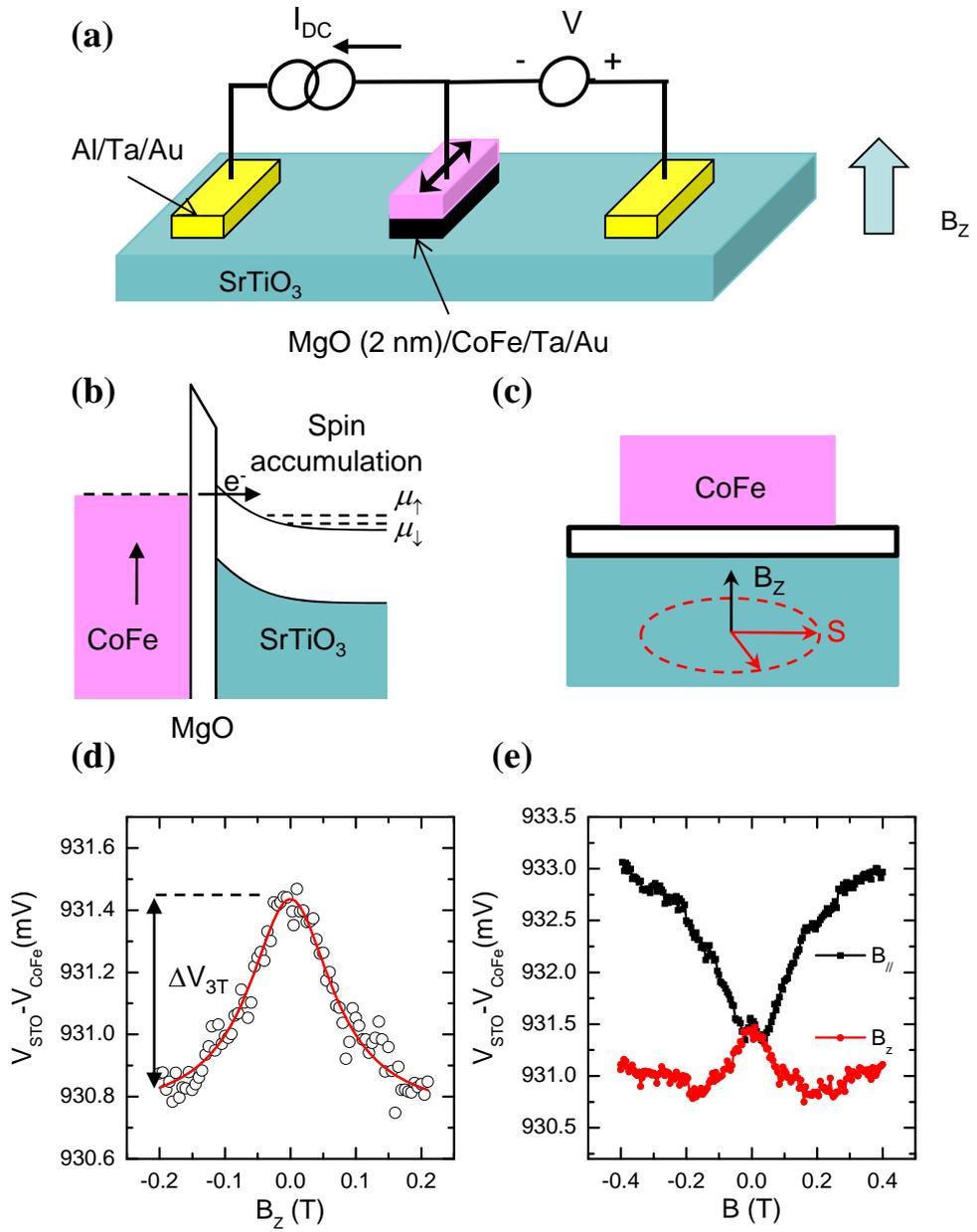

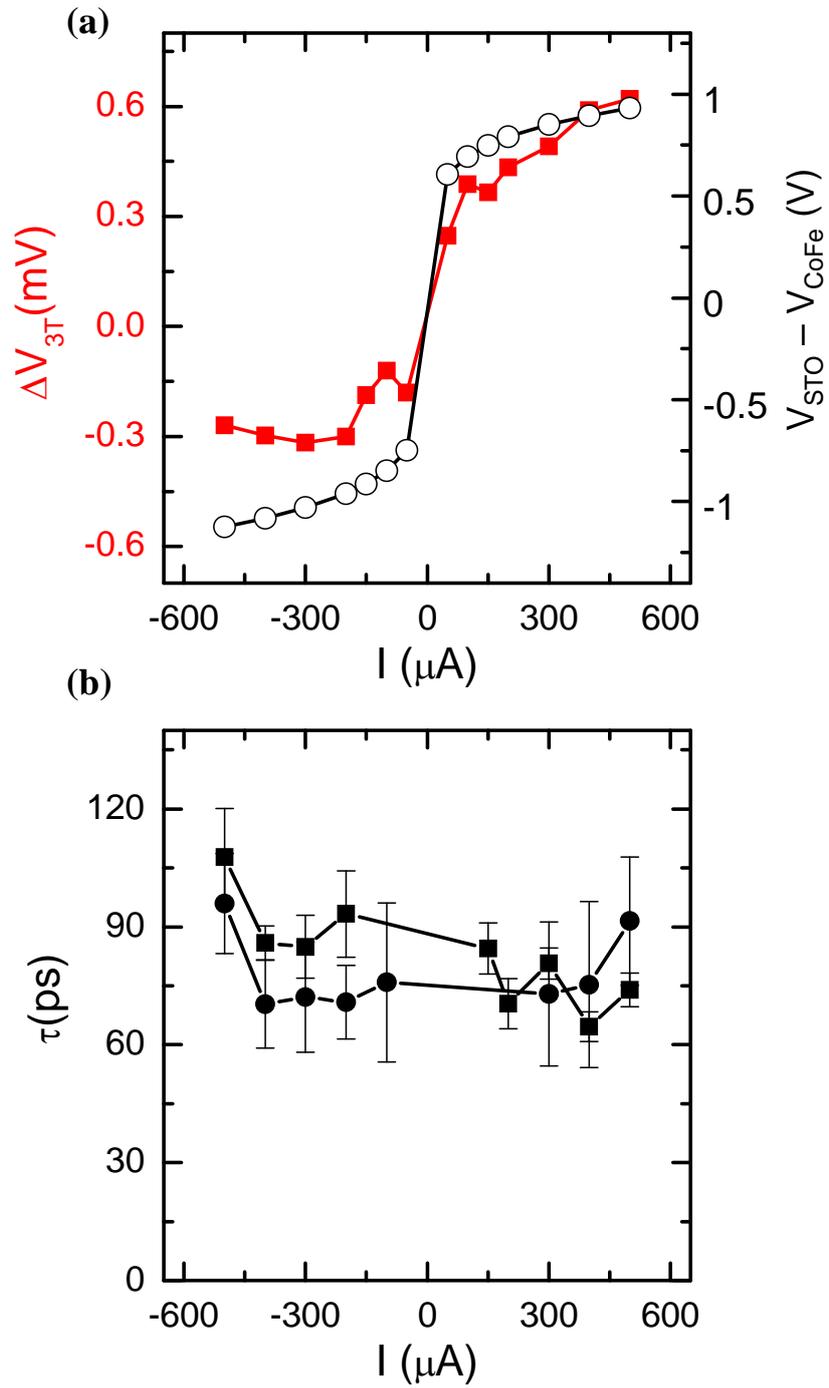

Figure 2

Figure 3

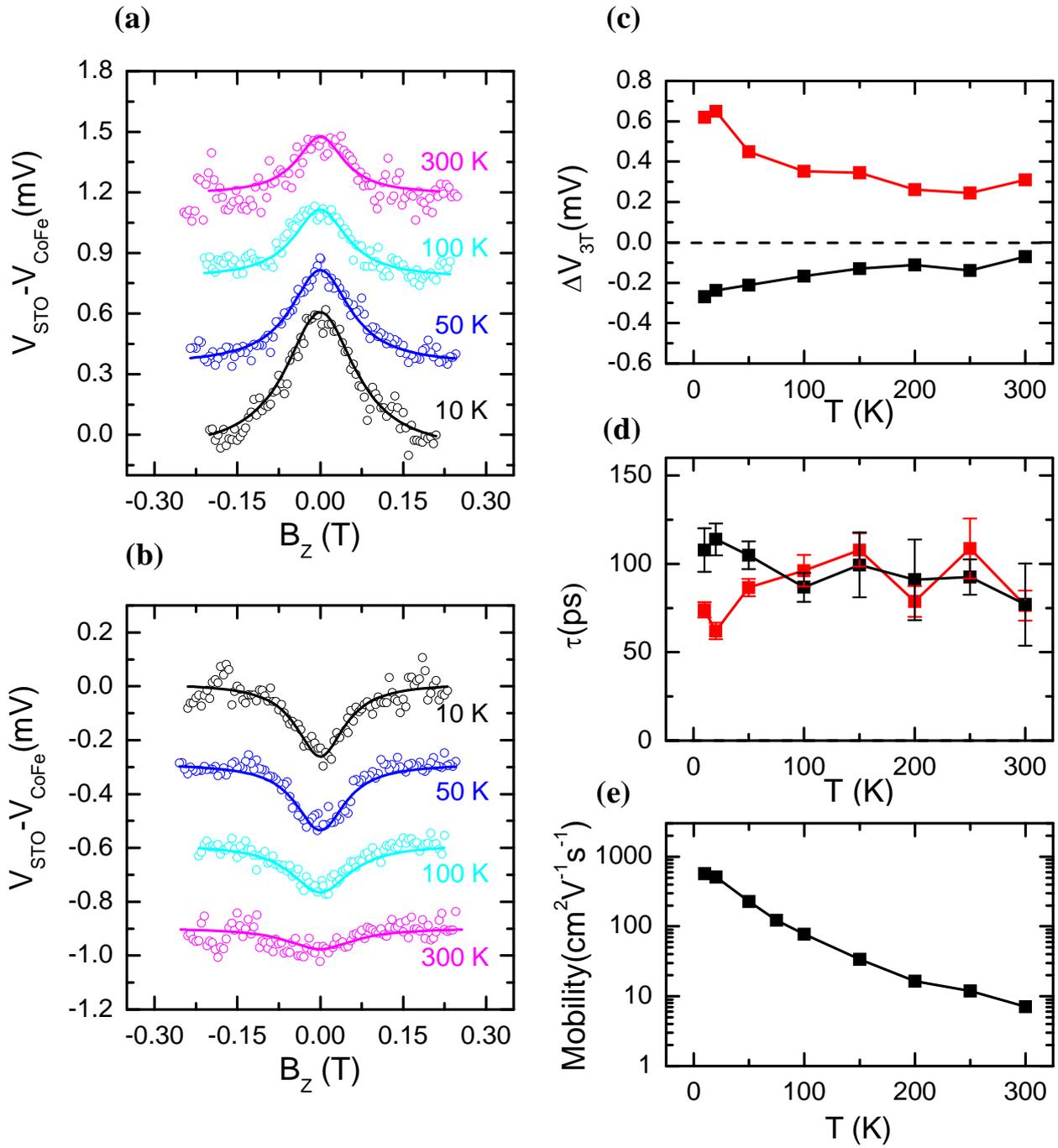

Figure 4

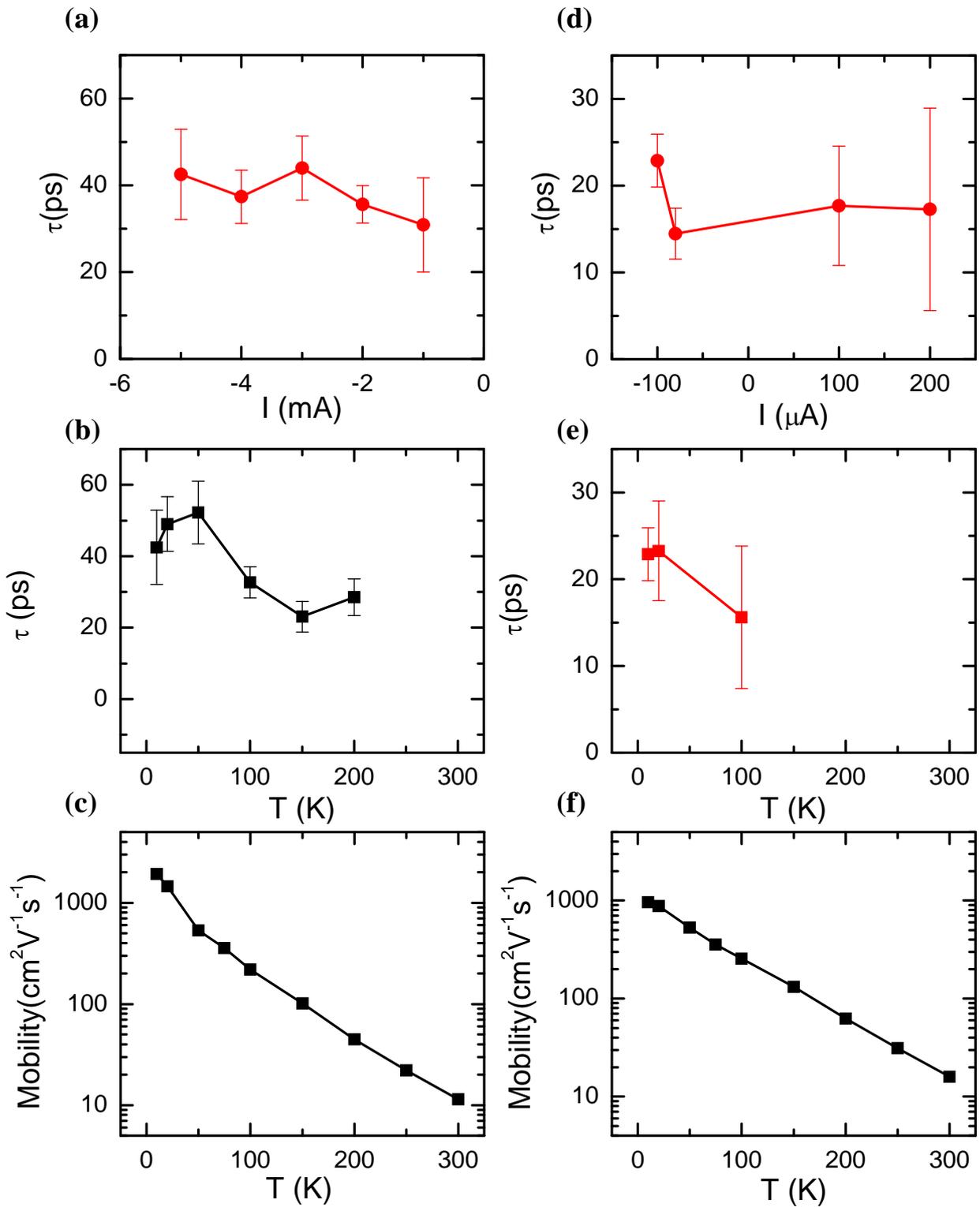

Figure 5

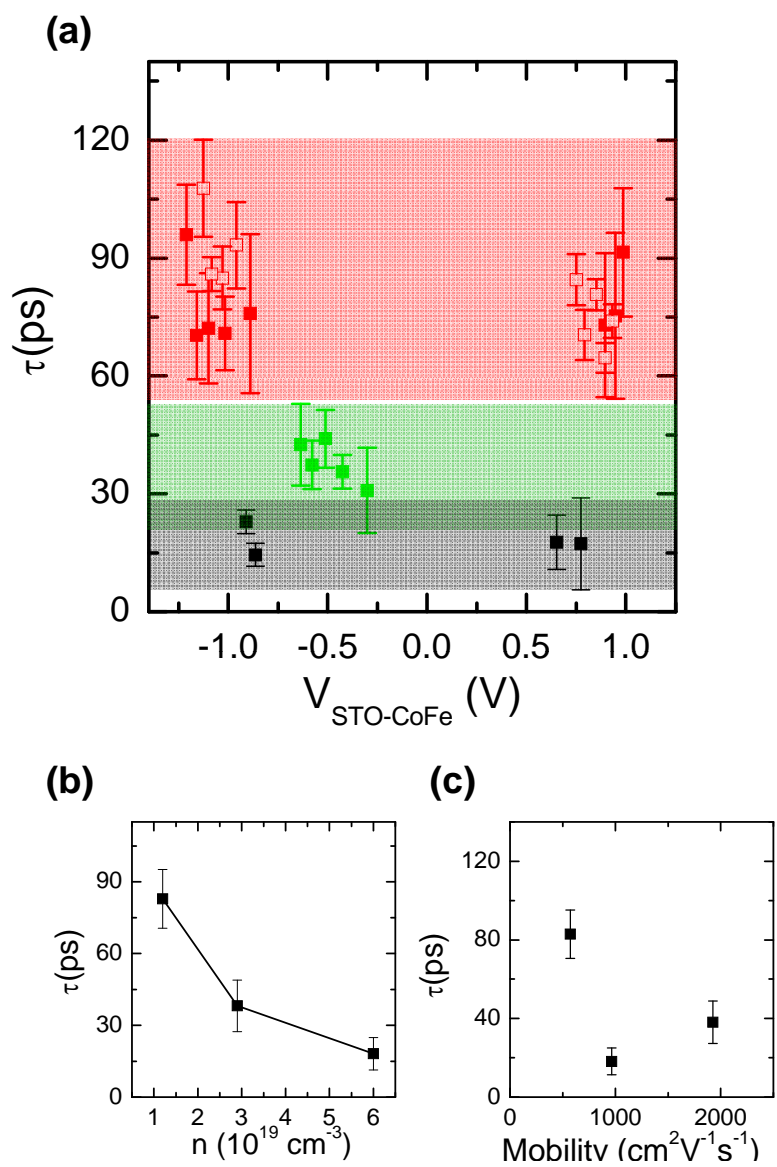

Figure 6

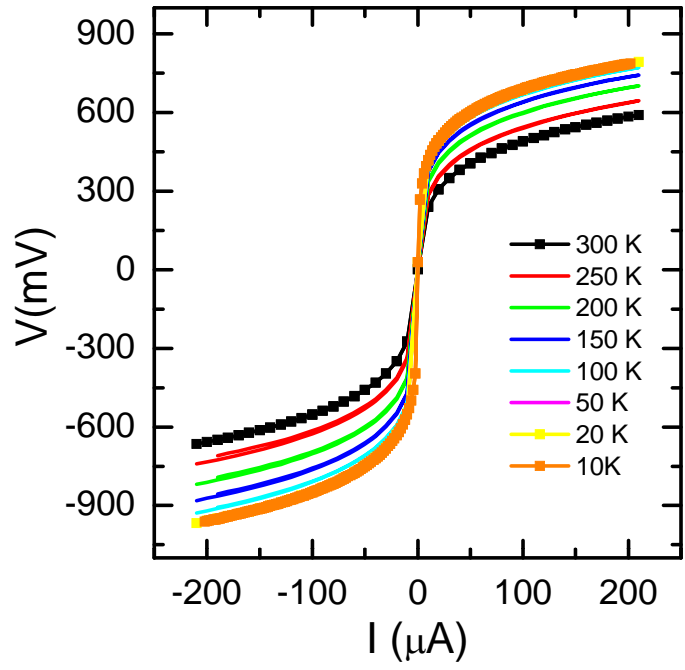